\documentstyle[preprint,aps]{revtex}
\begin{document}

\draft

\title{Conductance of a Finite Quantum Wire Connected to Reservoirs }

\author{Yu-Liang Liu}
\address{Max-Planck-Institut f\"{u}r Physik Komplexer Systeme, Bayreuther
Str. 40, D-01187 Dresden, Germany}

\maketitle

\begin{abstract}

We study a finite quantum wire connected to external leads, and show that the
conductance of the system significantly depends upon the length of the 
quantum wire and the position of the impurity in it. For a very long quantum
wire and the impurity far away from its two ends, the conductance has the same
behavior as that for an infinity quantum wire above some very little energy
scale. However, for a very short quantum wire, the conductance is independent
of the electron-electron interactions in it and closing to $e^{2}/(2\pi\hbar)$
in a higher temperature range. While, in a lower temperature range, the
conductance shows the same property as that for an infinity quantum wire.

\end{abstract}
\vspace{1cm}


\newpage

Recently, considerable efforts have been directed towards the study of the 
transport property of one-dimensional(1D) 
Tomonaga-Luttinger(TL) 
liquids\cite{1,2,3,4,5,6,7,8,9,10,11,12,13,14,15,16,17,18,19,19'}.
For an infinity impurity-free quantum wire, the conductance is believed to be
$ge^{2}/(2\pi\hbar)$ per spin orientation, where $g$ is a dimensionless
coupling strength parameter of the conduction electrons. For non-interacting
electrons, $g=1$. For repulsive interaction of the conduction electrons,
$g<1$, and the conductance is reduced. However, for a finite 
impurity-free quantum wire connected leads (reservoirs) which are
characterized by Fermi liquid, the conductance is believed to be
$e^{2}/(2\pi\hbar)$ which is independent of the interactions in the quantum
wire\cite{16,17,18,19,19'}, 
this surprising result derives from the boundary conditions
between the TL liquid in the wire and the Fermi liquid in the leads at the
ends of the wire. A recent experiment on a longer $GaAs$ high-mobility quantum
wires\cite{10} 
shows that for a higher temperature, the conductance is very close to
$e^{2}/(2\pi\hbar)$, but for a lower temperature, the conductance has a
power-law temperature dependence behavior, which is believed to be induced by
the impurity scattering in the quantum wires. Therefore, this experiment
really reveals the physical property of a finite quantum wire with 
impurity scattering. While, it is well-known that the backward scattering of
the conduction electrons induced by the impurity is relevant in terminology of
renormalization group, the perturbation methods may fail for treating this
kind of system, some results obtained by perturbation methods are not
reliable.  
In Ref.\cite{15}, by using bosonization method and unitary transformation,
we can exactly treat the backward scattering of the conduction electrons and
clearly show that the backward scattering significantly changes the
correlation exponents of the conduction electrons. It is very difficult to
obtain these correlation exponents by the perturbation methods. 
In present Letter, encouraged by the exact solution of single quantum impurity
scattering in TL-liquid, we study the conductance of a finite quantum wire 
connected to reservoirs. In contrast with an infinite quantum wire, 
it shows some different behaviors. The conductance drastically depends upon
the length of the quantum wire and the position of the impurity. For a very
long quantum wire, generally, it shows the same behavior as that for an
infinity quantum wire above some very little energy scale. For a very short
quantum wire, the conductance is independent of the electron-electron
interactions in the quantum wire in a higher temperature range, while its
temperature dependence has the same form as that for the infinity quantum wire
in a lower temperature range.

The Hamiltonian describing the 1D TL-liquid is generally given by 
\begin{equation}
H_{0}=-i\hbar v_{F}\int^{L-a}_{-a} dx[\psi^{+}_{R}(x)\partial_{x}\psi_{R}(x)
-\psi^{+}_{L}(x)\partial_{x}\psi_{L}(x)]
\label{1}\end{equation}
\begin{equation}
H_{I}=\frac{V}{2}\int^{L-a}_{-a} dx (\rho_{R}(x)+\rho_{L}(x))^{2}
\label{2}\end{equation}
\begin{equation}
H_{im}=V_{2k_{F}}[\psi^{+}_{R}(0)\psi_{L}(0)+\psi^{+}_{L}(0)\psi_{R}(0)]
\label{3}\end{equation}
where $\psi_{R}(x) (\psi^{+}_{R}(x))$ is the field operator of fermions 
that propagate to the right with wave vectors $\sim+k_{F}$, $\psi_{L}(x) 
(\psi^{+}_{L}(x))$ is the field operator of left propagating fermions 
with wave vectors $\sim -k_{F}$; 
$\rho_{R(L)}(x)=\psi^{+}_{R(L)}(x)\psi_{R(L)}(x)$ are the electron 
density operators; the spectrum of the electrons is linearized near the 
Fermi points and $v_{F}$ is the Fermi velocity; $V$ describes 
density-density interaction with momentum transferring much smaller than 
$k_{F}$.  
$V_{2k_{F}}=V(k=2k_{F})$ is the backward scattering potential of an impurity
residing at $x=0$ on the conduction electrons, for simplicity,
we have omitted the forward scattering potential because it has less
influence on the conductance. $L$ is the length of the quantum wire, and 
$0<a\leq L/2$. For simplicity, the reservoirs are assumed to be described by 
free electron systems. 

In the previous bosonization treatment of the Hamiltonians (\ref{1}), (\ref{2})
and (\ref{3}), one directly substitutes the bosonic representation of the
fermion fields $\psi_{R(L)}(x)$ into Equ.(\ref{3}) and obtains a
non-linear term which make the system be more difficult treated.
To more effectively study the physical property of the system described by the 
Hamiltonians (1),(2) and (3), we choose other new fermionic field operators
\begin{equation}
\psi_{1}(x)=\frac{1}{\sqrt{2}}(\psi_{R}(x)+\psi_{L}(-x)), \;\;\;
\psi_{2}(x)=\frac{1}{\sqrt{2}}(\psi_{R}(x)-\psi_{L}(-x))
\label{4}\end{equation}
It is easy to check that the operators $\psi_{1(2)}(x)$ satisfy the standard 
anticommutation relations. In terms of these new fermion fields
$\psi_{1(2)}(x)$, the cross term of the fermion fields
$\psi_{R(L)}(x)$ in (\ref{3}) can be written in a very simple form which
can be cancelled by a simple unitary transformation. However, the Hamiltonian
(\ref{2}) becomes complex.
Taking usual bosonization procedure\cite{1,20,21} for 
$\psi_{1(2)}(x)$, the Hamiltonians (1), (2) and (3) can be 
written as
\begin{equation}
H_{0}= 
\frac{\hbar v_{F}}{4\pi}\int^{L-a}_{-a} dx[(\partial_{x}\Phi_{1}(x))^{2}
+(\partial_{x}\Phi_{2}(x))^{2}]
\label{5}\end{equation}
\begin{eqnarray}
H_{I} &=& \displaystyle{\frac{V}{4}\int^{L-a}_{-a} 
dx\{[\rho_{1}(x)+\rho_{2}(x)]^{2}+
[\rho_{1}(x)+\rho_{2}(x)][\rho_{1}(-x)+\rho_{2}(-x)]} \nonumber \\
&+& \displaystyle{[\psi^{+}_{1}(x)\psi_{2}(x)+
\psi^{+}_{2}(x)\psi_{1}(x)]^{2}} \label{6} \\
&-& \displaystyle{[\psi^{+}_{1}(x)\psi_{2}(x)+
\psi^{+}_{2}(x)\psi_{1}(x)][\psi^{+}_{1}(-x)\psi_{2}(-x)+
\psi^{+}_{2}(-x)\psi_{1}(-x)]}\} 
\nonumber\end{eqnarray}
\begin{equation}
H_{im}=\frac{\hbar v_{F}\delta}{2\pi}(\partial_{x}\Phi_{1}(x)-
\partial_{x}\Phi_{2}(x))|_{x=0}
\label{7}\end{equation}
where $\delta=\arctan(V_{2k_{F}}/(\hbar v_{F}))$ is a phase shift induced by
the backward scattering potential $V_{2k_{F}}$. This replace of $V_{2k_{F}}$
by $\delta$ can be judged by the usual Born-approximation method and the
solution of the X-ray absorption in usual metals\cite{n}.
For simplicity, we write out the last two terms in (\ref{6}) by the fermion
fields $\psi_{1(2)}(x)$. If we use the boson fields $\Phi_{1(2)}$ to write out
these two terms, they become very complex $cosine$ forms. However, no matter
which description we use, the final result is same.
The bosonization representation of the fermion fields 
$\psi_{1(2)}$ can be written as\cite{1,20,21}
\begin{equation}
\psi_{1(2)}(x)=(\frac{D}{2\pi\hbar v_{F}})^{1/2}
\exp\{-i\Phi_{1(2)}(x)\}
\label{8}\end{equation}
where $D$ is the band width of the conduction electrons in the quantum wire,
$\rho_{1(2)}(x)=\psi^{+}_{1(2)}(x)\psi_{1(2)}(x)$ are the density 
operators of the fermion fields $\psi_{1(2)}(x)$, and have relations with the
boson fields $\Phi_{1(2)}(x)$:
$\partial_{x}\Phi_{1(2)}(x)=2\pi\rho_{1(2)}(x)$. In terms of the new boson and
fermion fields $\Phi_{1(2)}(x)$ and
$\psi_{1(2)}(x)$,  the Hamiltonian (\ref{6}) 
becomes complex, but the Hamiltonian (\ref{7}) becomes very simple. 
Generally, due to the simple form of the Hamiltonian (\ref{7}) which is
proportional to the density of the fermion fields $\psi_{1(2)}(x)$ at the
impurity site $x=0$, using an 
unitary transformation we can eliminate it, so that the problem is 
simplified as that we only need to treat a Hamiltonian similar to  
(\ref{5}) and (\ref{6}). However, the backward scattering interaction
drastically influences the behavior of the conduction electrons through
changing interactions among them. Therefore, we cannot simply eliminate the
backward scattering term by an unitary transformation and meanwhile leave the
Hamiltonians (\ref{5}) and (\ref{6}) intact.

To cancel the $\delta$-term in (\ref{7}) and simplify the system, 
we adopt the following steps\cite{15}:
\begin{description}
\item[1).] Taking the unitary transformation
\begin{equation}
U=\exp\{i\frac{\delta}{2\pi}
(\Phi_{1}(0)-\Phi_{2}(0))\}
\label{9}\end{equation}
we have the following relations
\[
U^{+}(H_{0}+H_{im})U=H_{0} \nonumber
\]
\begin{eqnarray}
U^{+}H_{I}U &=& \displaystyle{\frac{V}{4}\int^{L-a}_{-a} 
dx\{[\rho_{1}(x)+\rho_{2}(x)]^{2}+
[\rho_{1}(x)+\rho_{2}(x)][\rho_{1}(-x)+\rho_{2}(-x)]} \nonumber \\
&+& \displaystyle{[e^{-i\delta\;sgn(x)}\psi^{+}_{1}(x)\psi_{2}(x)+
e^{i\delta\;sgn(x)}\psi^{+}_{2}(x)\psi_{1}(x)]^{2}} \nonumber \\
&-& \displaystyle{[e^{-i\delta\;sgn(x)} \psi^{+}_{1}(x)\psi_{2}(x)+
e^{i\delta\;sgn(x)} \psi^{+}_{2}(x)\psi_{1}(x)]} \nonumber \\
&\cdot & \displaystyle{
[e^{i\delta\;sgn(x)} \psi^{+}_{1}(-x)\psi_{2}(-x)+
e^{-i\delta\;sgn(x)} \psi^{+}_{2}(-x)\psi_{1}(-x)]}\} 
\nonumber\end{eqnarray}
\item[2).] Performing the gauge transformations
\begin{equation}
\psi_{1(2)}(x)=\bar{\psi}_{1(2)}(x)e^{i\theta_{1(2)}}, \;\;\;
\theta_{1}-\theta_{2}=\delta
\label{10}\end{equation}
we have the relations
\[
U^{+}H_{I}U=\bar{H}^{(1)}_{I}+\bar{H}^{(2)}_{I}
\]
\begin{eqnarray}
\bar{H}^{(1)}_{I} &=& \displaystyle{\frac{V}{4}\int^{L-a}_{-a} 
dx\{[\rho_{1}(x)+\rho_{2}(x)]^{2}+
[\rho_{1}(x)+\rho_{2}(x)][\rho_{1}(-x)+\rho_{2}(-x)]} \nonumber \\
&+& \displaystyle{[\bar{\psi}^{+}_{1}(x)\bar{\psi}_{2}(x)+
\bar{\psi}^{+}_{2}(x)\bar{\psi}_{1}(x)]^{2}} \nonumber \\
&-& \displaystyle{\cos(2\delta)[\bar{\psi}^{+}_{1}(x)\bar{\psi}_{2}(x)+
\bar{\psi}^{+}_{2}(x)\bar{\psi}_{1}(x)]
[\bar{\psi}^{+}_{1}(-x)\bar{\psi}_{2}(-x)+
\bar{\psi}^{+}_{2}(-x)\bar{\psi}_{1}(-x)]}\} 
\nonumber\end{eqnarray}
\begin{eqnarray}
\bar{H}^{(2)}_{I} &=& \displaystyle{
\frac{V}{4}\int^{L-a}_{0} dx\{\frac{\cos(4\delta)-1}{2}[
\bar{\psi}^{+}_{1}(x)\bar{\psi}_{2}(x)+
\bar{\psi}^{+}_{2}(x)\bar{\psi}_{1}(x)]^{2} } \nonumber \\
&+& \displaystyle{[\bar{\psi}^{+}_{1}(x)\bar{\psi}_{2}(x)-
\bar{\psi}^{+}_{2}(x)\bar{\psi}_{1}(x)]^{2}\}} \nonumber \\
&+& \displaystyle{i\sin(2\delta)\int^{L-a}_{-a} dx
[\bar{\psi}^{+}_{1}(x)\bar{\psi}_{2}(x)-
\bar{\psi}^{+}_{2}(x)\bar{\psi}_{1}(x)]} \nonumber \\
&\cdot & \displaystyle{
[\bar{\psi}^{+}_{1}(-x)\bar{\psi}_{2}(-x)+
\bar{\psi}^{+}_{2}(-x)\bar{\psi}_{1}(-x)]}
\nonumber\end{eqnarray}
\item[3).] Re-defining the left- and right-moving electron fermions
\begin{eqnarray}
\bar{\psi}_{R}(x) &=& 
\displaystyle{\frac{1}{\sqrt{2}}[\bar{\psi}_{1}(x)+
\bar{\psi}_{2}(x)]}, \;\;\;
\bar{\psi}_{L}(-x)= 
\displaystyle{\frac{1}{\sqrt{2}}[\bar{\psi}_{1}(x)-
\bar{\psi}_{2}(x)]} \nonumber \\
\bar{\psi}_{R(L)}(x) &=& \displaystyle{
(\frac{D}{2\pi\hbar v_{F}})^{1/2}\exp\{-i\bar{\Phi}_{R(L)}(x)\}}, \;\;
\partial_{x}\bar{\Phi}_{R(L)}(x)=\pm 2\pi\bar{\rho}_{R(L)}(x)
\label{11}\end{eqnarray}
where $\bar{\rho}_{R(L)}(x)=\bar{\psi}^{+}_{R(L)}(x)\bar{\psi}_{R(L)}(x)$ are
the density operators of the electron fields $\bar{\psi}_{R(L)}(x)$,
the Hamiltonians $\bar{H}^{(1)}_{I}$ and $\bar{H}^{(2)}_{I}$ can be rewritten
as 
\begin{eqnarray}
\bar{H}^{(1)}_{I} &=& \displaystyle{ \frac{V}{2}\int^{L-a}_{-a} dx\{
[\bar{\rho}_{R}(x)+\bar{\rho}_{L}(x)]^{2}} \nonumber \\
&+& \displaystyle{\frac{1-\cos(2\delta)}{2}
[\bar{\rho}_{R}(x)-\bar{\rho}_{L}(-x)]
[\bar{\rho}_{R}(-x)-\bar{\rho}_{L}(x)]\}}
\nonumber \end{eqnarray}
\begin{eqnarray}
\bar{H}^{(2)}_{I} &=& \displaystyle{i\frac{V\sin(2\delta)}{4}\int^{L-a}_{-a}
dx [\bar{\rho}_{R}(-x)-\bar{\rho}_{L}(x)]
[\bar{\psi}^{+}_{L}(-x)\bar{\psi}_{R}(x)-
\bar{\psi}^{+}_{R}(x)\bar{\psi}_{L}(-x)]} \nonumber \\
&+& \displaystyle{\frac{V}{8}(1-\cos(4\delta))\int^{L-a}_{0} dx
[\bar{\rho}_{R}(-x)\bar{\rho}_{L}(x)-\bar{\rho}_{R}(x)\bar{\rho}_{L}(-x)]}
\nonumber\end{eqnarray}
It is worth notice that the Hamiltonian $\bar{H}^{(2)}_{I}$ only contributes
high order corrections because the first term has the conformal dimension
$\Delta\geq 2$ and the last term has the conformal dimension $\Delta=2$, and
at the weak ($\delta\sim 0$) and strong ($\delta\sim \pm\pi/2$) coupling
limit, they all tend to zero. Therefore, for simplicity we can neglect it.
\item[4).] Defining a set of new boson fields
\begin{eqnarray}
\Theta_{+}(x) &=& \displaystyle{
(\frac{G(\delta)}{\cosh(\chi_{1}-\chi_{2})})^{1/2}
[\cosh(\chi_{1})\Phi_{+}(x)-\sinh(\chi_{1})\Phi_{+}(-x)]} \nonumber \\
\Theta_{-}(x) &=& \displaystyle{
(\frac{1}{G(\delta)\cosh(\chi_{1}-\chi_{2})})^{1/2}
[\cosh(\chi_{2})\Phi_{-}(x)-\sinh(\chi_{2})\Phi_{-}(-x)]} 
\label{12}\end{eqnarray}
where $G(\delta)=[(1-\gamma)(1-\gamma\cos(2\delta))]^{1/4}/[(1+\gamma)
(1+\gamma\cos(2\delta))]^{1/4}$,
$\tan(2\chi_{1})=\beta\gamma/(1-\alpha\gamma)$, 
$\tan(2\chi_{2})=\beta\gamma/(1+\alpha\gamma)$, 
$\alpha=(1+\cos(2\delta))/2$, $\beta=(1-\cos(2\delta))/2$, $\Phi_{\pm}(x)=
[\bar{\Phi}_{R}(x)\pm\bar{\Phi}_{L}(x)]/\sqrt{2}$.
\end{description}
The total Hamiltonian of the system can be simplified as a very simple form
\begin{equation}
\bar{H}=\frac{\hbar\bar{v}_{F}}{4\pi g}\int^{L-a}_{-a} dx[
(\partial_{x}\Theta_{+}(x))^{2}+(\partial_{x}\Theta_{-}(x))^{2}]
\label{13}\end{equation}
where $\bar{v}_{F}=v_{F}\cosh(\chi_{1}-\chi_{2})(
\frac{1-(\gamma\cos(2\delta))^{2}}{1-\gamma^{2}})^{1/4}$,
$g=(\frac{1-\gamma}{1+\gamma})^{1/2}$ is a dimensionless coupling
strength parameter, where $\gamma=V/(2\pi\hbar v_{F}+V)$. 
Here we have omitted some higher order terms\cite{15}
which only give less important high order correction to the conductance. 
However, the dual boson fields $\Theta_{\pm}(x)$ satisfy
the following commutation relations
\begin{eqnarray}
[\partial_{x}\Theta_{+}(x), \;\; \Theta_{-}(y)] &=&
i2\pi\delta(x-y)+i2\pi\delta(x+y)\tan(\chi_{1}-\chi_{2}) \nonumber \\
\displaystyle{
[\partial_{x}\Theta_{-}(x), \;\; \Theta_{+}(y)]} &=&
i2\pi\delta(x-y)+i2\pi\delta(x+y)\tan(\chi_{1}-\chi_{2}) 
\label{14}\end{eqnarray}
which have an anomaly term $i2\pi\delta(x+y)\tan(\chi_{1}-\chi_{2})$. This term
is zero at the weak $\delta\sim 0$ and strong $\delta\sim\pm\pi/2$ coupling
limits. According to these commutation relations of the dual boson fields
$\Theta_{\pm}(x)$, for example, we can define the conjugate momentum field
$P_{-}(x)$ of the
boson field $\Theta_{-}(x)$ as
\begin{equation}
-\frac{1}{2\pi}\partial_{x}\Theta_{+}(x)=
P_{-}(x)+\tan(\chi_{1}-\chi_{2})P_{-}(-x),\;\;
[P_{-}(x),\; \Theta_{-}(y)]=-i\delta(x-y)
\label{15}\end{equation}
and then we can exactly solve the Hamiltonian (\ref{13}). Therefore, we can
calculate the conductance for any backward
scattering potential.

For simplicity, we first consider a special case: 
$\delta=0$, which corresponds to the weak backward
scattering limits. In this case, the propagators of the boson
field $\Theta_{-}(x)$ satisfy the following equation
\begin{equation}
\{-\partial_{x}(\frac{v}{g}\partial_{x})+\frac{\tilde{\omega}^{2}}{vg}
\}G^{-}_{\tilde{\omega}}(x,x')=\delta(x-x')
\label{16}\end{equation}
where $G^{-}_{\tilde{\omega}}(x,x')=\frac{1}{2\pi}\int^{1/(k_{B}T)}_{0}
d\tau <T_{\tau}\Theta_{-}(x,\tau)\Theta_{-}(x',0)>e^{i\tilde{\omega}
\tau}$, where $T$ is temperature, $v=v_{F}/g$. 
Using the boundary conditions\cite{16} of 
the propagator $G^{-}_{\tilde{\omega}}(x,x')$, we can easily obtain the
following equation at the impurity site $x=0$
\begin{equation}
G^{-}_{\tilde{\omega}}(0,0)\simeq\frac{K}{2|\tilde{\omega}|}
\label{17}\end{equation}
where the dimensionless coupling strength parameter $K$ satisfies the
following relations
\begin{equation}
K=\left\{\begin{array}{ll}
g, & \;\; |\tilde{\omega}|\gg v/a\\
\displaystyle{\frac{2g}{g+1}}, & \;\; v/L\ll|\tilde{\omega}|\ll v/a\\
1, & \;\; |\tilde{\omega}|\ll v/L
\end{array}\right.
\label{18}\end{equation} 
However, for a general phase shift $\delta$, we can
obtain the following expression of the propagator 
$G^{-}_{\tilde{\omega}}(x,x')$ at the impurity site $x=0$
\begin{equation}
G^{-}_{\tilde{\omega}}(0,0)\simeq\frac{G(\delta)\bar{K}}
{2\Gamma |\tilde{\omega}|}
\label{19}\end{equation}
where the dimensionless coupling strength parameter $\bar{K}$ satisfies the
following relations
\begin{equation}
\bar{K}=\left\{\begin{array}{ll}
1, & \;\; |\tilde{\omega}|\gg v^{'}/a\\
\displaystyle{\frac{2\Gamma}{\Gamma+G(\delta)}}, & \;\; 
v^{'}/L\ll|\tilde{\omega}|\ll v^{'}/a\\
\displaystyle{\frac{\Gamma}{G(\delta)}}, & \;\; |\tilde{\omega}|\ll v^{'}/L
\end{array}\right.
\label{20}\end{equation}
where $\Gamma=[1+\tan(\chi_{1}-\chi_{2})]/[1+\tan^{2}(\chi_{1}-\chi_{2})]$,
$v^{'}=\bar{v}_{F}/g$. 
 
Now we calculate the conductance of the system. To this end, we first define a
charge density operator $Q(x)$, and then use the continuous equation to obtain
the current density operator $J(x)$. The charge density operator $Q(x)$ is
equal to $e(\rho_{R}(x)+\rho_{L}(x))$. Under the unitary and gauge 
transformations (\ref{9}) and (\ref{10}), it can be written as
\begin{equation}
U^{+}\rho_{R}(x)U=\left\{ \begin{array}{ll}
& \displaystyle{ \bar{\rho}_{R}(x)-\beta
(\bar{\rho}_{R}(x)-\bar{\rho}_{L}(-x))}+\\ 
& -i\sin(2\delta)(\bar{\psi}^{+}_{L}(-x)\bar{\psi}_{R}(x)-
\bar{\psi}^{+}_{R}(x)\bar{\psi}_{L}(-x)), \;\;\;\; x>0\\
& \bar{\rho}_{R}(x), \;\;\;\;\;\; x<0
\end{array}\right.
\label{21}\end{equation}
\begin{equation}
U^{+}\rho_{L}(x)U=\left\{ \begin{array}{ll}
& \bar{\rho}_{L}(x), \;\;\;\;\;\; x>0\\
&\displaystyle{ \bar{\rho}_{L}(x)+\beta
(\bar{\rho}_{R}(-x)-\bar{\rho}_{L}(x))}+\\ 
& -i\sin(2\delta)(\bar{\psi}^{+}_{L}(-x)\bar{\psi}_{R}(x)-
\bar{\psi}^{+}_{R}(x)\bar{\psi}_{L}(-x)), \;\;\;\; x<0
\end{array}\right.
\label{22}\end{equation}
It is worth noting that if the phase shift $\delta$ takes the values
$\pm\pi/2$, the electrons are completely reflected at the impurity site $x=0$,
and we have the relations: $U^{+}\rho_{R}(x)U=\bar{\rho}_{L}(-x)$ for $x>0$,
and $U^{+}\rho_{L}(x)U=\bar{\rho}_{R}(-x)$ for $x<0$. Therefore, the phase
shift $\delta^{c}=\pm\pi/2$ correspond to the strong coupling critical points
of the system. However, there exists a gauge symmetry in the system, if we
take $\theta_{1}-\theta_{2}=-\delta$, we can have the relations at the strong
coupling critical points: $U^{+}\rho_{R}(x)U=\bar{\rho}_{L}(-x)$ for $x<0$,
and $U^{+}\rho_{L}(x)U=\bar{\rho}_{R}(-x)$ for $x>0$.
Accordingly, for $x>0$, we can obtain the following current density operator
\begin{eqnarray}
J(x) &=& \displaystyle{
\frac{e}{2\pi}(\frac{\cosh(\chi_{1}-\chi_{2})}{2})^{1/2}\{-\beta(\frac{
g_{1}(\delta)}{G(\delta)})^{1/2}[\partial_{\tau}\Theta_{+}(x)-
\partial_{\tau}\Theta_{+}(-x)]} \nonumber \\
&+& \displaystyle{
(g_{2}(\delta)G(\delta))^{1/2}[\partial_{\tau}\Theta_{-}(x)-
\partial_{\tau}\Theta_{-}(-x)]} \nonumber \\
&+& \displaystyle{
(\frac{G(\delta)}{g_{2}(\delta)})^{1/2}[\partial_{\tau}\Theta_{-}(x)+
\partial_{\tau}\Theta_{-}(-x)]\}} \nonumber \\
&-& \displaystyle{i\sin(2\delta)\int^{x}_{0} dy\partial_{\tau}
[\bar{\psi}^{+}_{L}(-y)\bar{\psi}_{R}(y)-
\bar{\psi}^{+}_{R}(y)\bar{\psi}_{L}(-y)]}
\label{23}\end{eqnarray}
where $g_{1}(\delta)=(1-\gamma)^{1/2}/(1-\gamma\cos(2\delta))^{1/2}$,
$g_{2}(\delta)=(1+\gamma\cos(2\delta))^{1/2}/(1+\gamma)^{1/2}$.
Using Kubo formula of the conductance, and the expression of the propagators
$G^{-}_{\tilde{\omega}}(x\sim 0, x'\sim 0)$ in (\ref{19}) and (\ref{20}), we
can obtain the following electric conductance
\begin{equation}
\sigma_{\tilde{\omega}}(x\sim x'\sim 0)=\left\{\begin{array}{ll}
\displaystyle{ \frac{e^{2}}{2\pi\hbar}\frac{\alpha^{2}G(\delta)\cosh(\chi_{1}-
\chi_{2})}{g_{2}(\delta)\Gamma}+A\tilde{\omega}^{2\mu}}, & \;\;\;
|\tilde{\omega}|\gg v^{'}/a\\
\displaystyle{ \frac{e^{2}}{2\pi\hbar}\frac{\alpha^{2}\cosh(\chi_{1}-
\chi_{2})}{g_{2}(\delta)}+B\tilde{\omega}^{2\nu}}, & \;\;\;
|\tilde{\omega}|\ll v^{'}/L
\end{array}\right.
\label{24}\end{equation}
where $\mu=G(\delta)\cosh(\chi_{1}-\chi_{2})/(g_{2}(\delta)\Gamma)$, 
$\nu=\cosh(\chi_{1}-\chi_{2})/g_{2}(\delta)$,
$A$ and $B$ are constants. This is our central result of present paper. The
temperature dependence of the electric conductance
$\sigma_{\tilde{\omega}}(x\sim x'\sim 0)$ can be obtained through replacing the
frequency $\tilde{\omega}$ by the temperature $T$. It is necessary
to mention that as the frequency $\tilde{\omega}$ and the temperature $T$ tend
to zero, $\{\bar{\omega},\;T\}\rightarrow 0$, the phase shift $\delta$ takes
the values $\pm\pi/2$, and the electric conductance is equal to zero. This
behavior can be easily understood by using the renormalization group\cite{5}
that because the backward scattering term is relevant, the renormalized
backward scattering potential $\bar{V}_{2k_{F}}$ goes to infinity in the low
energy limit, therefore, the phase shift $\delta$ induced by the backward
scattering potential takes the values $\pm\pi/2$. The electric conductance
(\ref{24}) does not contradict previous work. As
$\{\bar{\omega},\;T\}= 0$, the electrons are completely reflected on the
impurity site $x=0$ for the repulsive electron-electron interaction. 
It is worth notice that the electric conductance significantly depends upon the
length of the wire, and the position of the impurity. In
generally, for a very long wire, $L\rightarrow\infty$, the electric 
conductance is in
the range of $|\tilde{\omega}|\gg v^{'}/a$. For an impurity-free system,
$\delta=0$, the electric conductance is equal to $ge^{2}/(2\pi\hbar)$.   
If there exists impurity scattering, the electric conductance is
$\alpha^{2}e^{2}/(2\pi g\hbar )$ in the low energy limit. 
All these properties are
the same as that for an infinite quantum wire. However, for a short quantum
wire, in the low energy limit, the electric conductance falls in the range of 
$|\tilde{\omega}|\ll v^{'}/L$, for the impurity-free case, the electric 
conductance is
equal to $e^{2}/(2\pi\hbar)$ which is independent of the electron-electron
interactions in the quantum wire\cite{16,17,18,19,19'}. As including the
impurity scattering, for the weak backward scattering, the electric 
conductance is
still very closing to $e^{2}/(2\pi\hbar)$. For the strong backward scattering,
the electric conductance is the same as that for the infinite quantum wire. 
However, to
consider the temperature dependence of the conductance, 
we must calculate the
tunneling conductance at the impurity site $x=0$ which derives from the quantum
fluctuation of collective excitation modes of the system, 
because Eq.(\ref{24}) only gives a higher order temperature dependence.  

To calculate the tunneling conductance of the system, we can
define the following tunneling current operator at the impurity site $x=0$ as
\[
I_{tunn}=\frac{et_{0}}{2}(\psi^{+}_{R}(0)
\psi_{L}(0)+\psi^{+}_{L}(0)\psi_{R}(0))
\]
where $t_{0}$ is the tunneling probability amplitude. This definition and the
following calculation of the tunneling conductance are meaningful only at the
strong coupling region determined by the phase shift $\delta$.
In terms of the boson fields $\Theta_{\pm}(x)$, it can be written as
\begin{equation}
I_{tunn}=\frac{et_{0}D}{2\pi\hbar v_{F}}\cos[(\frac{2G(\delta)
\cosh(\chi_{1}-\chi_{2})}{g_{2}(\delta)})^{1/2}\Theta_{-}(0)]
\label{25}\end{equation}
Therefore, by using Kubo formula of the conductance, we can obtain the
following tunneling conductance 
\begin{equation}
\bar{\sigma}(\tilde{\omega})\sim\left\{\begin{array}{ll}
\displaystyle{\tilde{\omega}^{2(\mu-1)}}, 
& \;\;\; |\tilde{\omega}|\gg v^{'}/a\\
\displaystyle{\tilde{\omega}^{2(\nu-1)}}, & \;\;\; |\tilde{\omega}|\ll v^{'}/L
\end{array}\right.
\label{26}\end{equation}
It is worth notice that in the weak coupling fixed point, $\delta=0$, 
the exponents $\mu$ and $\nu$ take the values: $\mu=g$, and $\nu=1$. 
While in the strong coupling critical points, $\delta^{c}=\pm\pi/2$,  
the exponents $\mu$ and $\nu$ take the values: $\mu=\nu=1/g$. 
However, in the range of $v^{'}/L\ll |\tilde{\omega}|\ll v^{'}/a$, the
exponents $\mu$ and $\nu$ also depend upon the parameters $a$ and $L$. In
generally, it is difficult to compare with the experimental data in
Ref.\cite{10}. Based upon Eqs.(\ref{24}) and (\ref{26}), we can unambiguously
obtain the following conclusions
that for a very long quantum wire, $L\rightarrow\infty$, the usual
experimental energy falls in the range of $|\tilde{\omega}|\gg v^{'}/L$, if
the impurity site is far away from the ends of the quantum wire, the
conductance of the system shows the same behavior as that for an infinite
quantum wire. While, for the case of a short quantum wire, the usual
experimental energy is in the range of $|\tilde{\omega}|\ll v^{'}/L$, the
conductance is independent of the electron-electron interactions in the
quantum wire in a higher temperature range (but still leaving the condition 
$|\tilde{\omega}|\ll v^{'}/L$ intact), and its temperature dependence is the
same as that for an infinite quantum wire in the lower temperature range.

In summary, using the bosonization method and the unitary transformation, we
have studied the system of a finite quantum wire connected to leads, and shown
that in contrast with an infinity quantum wire, it significantly relies upon
the length of the wire and the position of the impurity in the wire. 
For a very short quantum wire, in the higher temperature range, the
conductance is independent of the electron-electron interactions in the wire
and is very closing to $e^{2}/(2\pi\hbar)$. However, in the lower temperature
range, it is the same as that for an infinity quantum wire.

We are very grateful to Prof. Peter Fulde for his 
encouragement.

\end{document}